\documentclass{PoS}
\usepackage{wrapfig}

\title{The nonperturbative phase diagram of the bosonic BMN matrix model}

\ShortTitle{Bosonic BMN phase diagram}

\author{\speaker{Samuel Kov\'a\v{c}ik} \\
 School of Theoretical Physics, Dublin Institute for Advanced Studies, 10 Burlington Road,  Dublin 4, Ireland.\\
        E-mail: \email{skovacik@stp.dias.ie}}

\author{Denjoe O'Connor\\
 School of Theoretical Physics, Dublin Institute for Advanced Studies, 10 Burlington Road,  Dublin 4, Ireland. \\
        E-mail: \email{denjoe@stp.dias.ie}}
        
        \author{Yuhma Asano\\
KEK Theory Center, High Energy Accelerator Research Organization,
1-1 Oho, Tsukuba, Ibaraki 305-0801, Japan.\\
        E-mail: \email{yuhma@post.kek.jp}}

\abstract{We study the thermal phase transition of the bosonic BMN model which is a mass deformed version of the bosonic part of the BFSS model. Our results connect the massless region of the phase diagram described by the bosonic BFSS model with the large-mass region, where the model is analytically solvable. We observe that at finite value of the matrix size $N$, the critical region is smeared over a small temperature range. The model has a single critical temperature, which arises as the large $N$ limit of two apparent transitions at finite $N$. We emphasise the vital role played by finite $N$ corrections in the confined phase and illustrate this with a novel treatment of the noninteracting Gaussian model.
\\ 
\\
Preprint numbers: DIAS-STP-20-06, KEK-TH-2210}

\FullConference{Corfu Summer Institute 2019 "School and Workshops on Elementary Particle Physics and Gravity" (CORFU2019)\\
		31 August - 25 September 2019\\
		Corfù, Greece}

\begin{document}

\section{Introduction}
Dimensionally reduced Yang-Mills models are among the best candidates for testing the gauge/ gravity conjecture, one of the most researched examples of which is the BFSS model \cite{deWit:1988wri,Banks:1996vh} and its maximally supersymmetric mass deformed version, the BMN model \cite{Berenstein:2002jq}. Bosonic and supersymetric models also arise from quantization of membranes and supermembranes on various backgrounds \cite{deWit:1988wri,Kim:2006wg}.

This family of quantum matrix models, which can also be interpreted as models of interacting D0-branes, has a surprisingly rich phase
structure including deconfining phase transitions as the temperature is varied
\cite{Aharony:2003sx,Furuuchi:2003sy,Semenoff:2005ei,Aharony:2005ew,Asano:2018nol,Asano:2020yry}.

In the large-mass limit, the BMN model reduces to a gauged Gaussian model that can be solved analytically and has a single phase transition. At zero mass, based on the gauge/gravity duality conjecture, the model is connected to the Gregory-Laflamme \cite{Gregory:1993vy,Gregory:1994bj} transition. Our goal is to connect those two regimes nonperturbatively using numerical simulations.

The massless version of the model, the bosonic BFSS model, has been studied extensively both analytically and numerically \cite{Catterall:2010gf, Aharony:2004ig,Kawahara:2007fn,Filev:2015hia,Azuma:2014cfa,Mandal:2009vz,Bergner:2019rca,Schaich:2020ubh}. Initial studies reported two close thermal phase transitions which were in a good agreement with the results from $1/D$ ($D$ is the number of matrices) expansion performed in \cite{Kawahara:2007fn}. Later it was realised that in the large-$N$ limit there is only a single phase transition \cite{Bergner:2019rca}, which appears to be of the 1st order. Our study of the bosonic BMN model for $\mu=2$ agrees with this conclusion \cite{Asano:2020yry}.

The massless bosonic BFSS model has received much attention already in previous studies \cite{Aharony:2004ig,Kawahara:2007fn,Filev:2015hia,Azuma:2014cfa,Mandal:2009vz,Bergner:2019rca,Morita:2020liy}.
The initial work reported only a single transition
\cite{Aharony:2004ig} but the $1/D$ expansion \cite{Mandal:2009vz} suggested existence of two closely separated transitions. This was supported by numerical studies \cite{Kawahara:2007fn} at small $N$. Later, a recent study \cite{Bergner:2019rca} at larger $N$ and new analytic results \cite{Morita:2020liy} find evidence of a single (first order) confining/deconfining phase transition. Our study \cite{Asano:2020yry} of the BMN model gives the same conclusion as \cite{Bergner:2019rca, Morita:2020liy} and reports only a single transition.

In this paper we report our findings regarding the phase structure of the bosonic BMN model. At finite $N$, we observed two distinct phase transitions which merge in the large-$N$ limit into a single one. Even though one of our approaches indicates that the transition is a standard 1st order one, showing clear signs of a transitioning two-level system, another approach suggests the transition might be more related to the Hagedorn transition. We leave this question open at this moment. 

For $\mu = 2 $ we gather enough data to extrapolate the results to the large-$N$ limit. For other values of $\mu$ we fixed $N=12$ and produced a phase diagram with two (pseudo)critical temperatures for each value of $\mu$. These are expected to merge in the large-$N$ limit and their finite-$N$ values can serve as upper and lower boundaries for the large-$N$ critical temperature. 

\section{Model and observables}

The gauged quantum model is defined using $D=9$ Hermitian $N \times N$ matrices that transform as adjoint representation of $SU(N)$ and are placed on a thermal circle with the action defined as

\begin{eqnarray}\label{BMNaction}
 S[X,A]
 =N\int \limits_0^{\beta} d \tau \ \, 
 \mbox{Tr } \Bigg[ &
 \frac{1}{2}D_\tau X^i D_\tau X^i 
 -\frac{1}{4} \left( [X^r,X^s]+\frac{i\mu}{3}\varepsilon^{rst}X_t \right)^2
 \\
  &
 -\frac{1}{2} [X^r,X^m]^2
 -\frac{1}{4} [X^m,X^n]^2
 +\frac{1}{2}\left( \frac{\mu}{6}\right)^2X_m^2
 \nonumber
 \Bigg] ,
\end{eqnarray}  
where $i=1,\dots,9$; $r,s=1, 2, 3$ and $m,n=4,\dots , 8, 9$. The mass parameter is $\mu$, $\beta = 1/T$ is the inverse temperature and $D_\tau\cdot=\partial_\tau\cdot-i[A,\cdot]$ is the covariant derivative. The $SO(9)$ symmetry is explicitly broken to $SO(6)\times SO(3)$
by the mass terms and the cubic Myers term. We fixed $A$ to be diagonal and time independent which invokes the Vandermonde determinant described in \cite{Filev:2015hia}.

Mean values of observables $\mathcal{O}$ are defined by path integration over Hermitian matrix elements as 
\begin{eqnarray}
\left\langle \mathcal{O} \right\rangle = \frac{\int [dX][dA] \ \mathcal{O} \ e^{-S[X,A]}}{Z}, \ Z = \int  [dX][dA] e^{-S[X,A]}.
\end{eqnarray}

We employ the usual lattice formulation where the matrices $X^i$ are placed on temporal sites and $A$ on the links between them. The (Euclidean) time $\tau$ is discretised as $\tau \rightarrow \beta k / \Lambda$, where $k = 1,\dots , \Lambda$. To reduce the discretisation effects from the kinetic term we use the method discussed in \cite{Asano:2018nol, Asano:2019pre}. The coupling constant has been fixed to $1$ and all
dimensional quantities are expressed in these natural units.

The standard set of observables for analysis of thermal phase transitions is the energy $E$, the specific heat $C_{\rm v}$, the extent of eigenvalues $\langle R^2\rangle $ and the Polyakov loop $\langle |P|\rangle$ which serves as an order parameter in the deconfining transition. The Myers observable, $M$, is important in the supersymmetric formulation of the model as fermionic degrees of freedom can stabilise fuzzy-sphere configurations  \cite{Asano:2018nol}, we have not observed such behaviour in the bosonic model. The list of observables follows: 
\begin{eqnarray} \nonumber
E &=&  N^{-2} (-\partial_\beta) \log Z , \\ \nonumber
C_{\rm v} &=&  \beta^2 \partial_\beta^2 \log Z , \\
\label{obs}
\langle |P|\rangle  &=& \left\langle \frac{1}{N}\left| \mbox{Tr } \left( \exp \left( i \beta A \right) \right)\right| \right\rangle , \\ \nonumber
\langle R^2\rangle &=& \left\langle \frac{1}{N \beta}  \int \limits_0^\beta d \tau \ \mbox{Tr }\left( X^i X^i \right) \right\rangle , \\ \nonumber
M &=&  \left\langle \frac{i}{3 N \beta}  \int \limits_0^\beta d \tau \ \varepsilon^{rst}  \mbox{ Tr }\left( X^r X^s X^t \right) \right\rangle .
\end{eqnarray}
Typical behaviour of these observables is shown in the figures \ref{Fig:Observables} and \ref{Fig:pols}.

We have also introduced two new observables that improve the accuracy of measurements of (pseudo)critical temperatures for finite values of $N$. We performed Hybrid Monte Carlo (HMC) simulations of the system to evaluate the path integrals and noticed that close to the apparent transition temperature, the system transitions between two distinct levels: one close to $\langle\vert
P\vert\rangle \approx 1/N$ and the other close to $\langle\vert
P\vert\rangle \approx 1/2$. At low temperature, the system spends the entire Monte Carlo time at the bottom level, but as the temperature is increased it tends to spend larger portion of it in the top one. Therefore, we defined the observable $\mathbb{P}$ that captures which level is preferred by the system at a given temperature defined by
\begin{equation}
  \mathbb{P}_x = \int \limits_{x}^1 \mathcal{P} (q) dq\, ,\quad\hbox{with}\quad
  \mathbb{P}_{0.5}=\mathbb{P},
\end{equation}
where $\mathcal{P}(q)$ is the probability distribution for the Polyakov loop. Quite surprisingly, this observable shows a very clear piecewise linear behaviour, see the figure \ref{Fig:Q} \footnote{We have tested that using $ \mathbb{P}_{0.4}$ or $ \mathbb{P}_{0.3}$ yields similar results.}. It is constant well bellow and above the transition and linear in the middle of it. We take the root of the linear function describing the transition region to be $T_{c1}$. As the steepness of this line increases with $N$ indefinitely, values of $T_{c1}$ defined by any point on it converge to the same value. 

The second introduced observable is 
\begin{equation}
\langle\vert P_n\vert\rangle = \left\langle \frac{1}{N}\left|  \mbox{Tr } \left( \exp \left( i n \beta A \right) \right)\right| \right\rangle .
\label{P_n}
\end{equation}
This observable for $n>2$ captures the behaviour of higher moments of the eigenvalue distribution of $A$ expressed as $u_n = \int \limits _{-\pi}^\pi\rho(\theta) e^{i n \theta} d \theta$. In the zero-temperature limit, the eigenvalues of $A$ are distributed uniformly, $u_n=0$ for $n \ge 1$.  Then, with increasing temperature, the distribution becomes nonuniform, $u_1>0$ while $u_2 = u_3 = ... = 0$. With further increasing temperature, the distribution develops a gap, all moments become excited or equivalently $\langle\vert P_n\vert\rangle > 0$ for $n\ge 1 $. For all values of $N$ we observe a very sharp change in the behaviour of $\langle\vert P_2\vert\rangle$. It is constant below a certain temperature and then starts growing above it. We denote this temperature by $T_{c2}$. Higher modes $\langle\vert P_n\vert\rangle, \ n = 3,4, ...$ are growing as well, but at a slower rate than $\langle\vert P_2\vert\rangle$ so we use it to mark the transition.

\section{Thermal phase transition(s)}

\begin{figure}[t] 
  \includegraphics[width=6in]{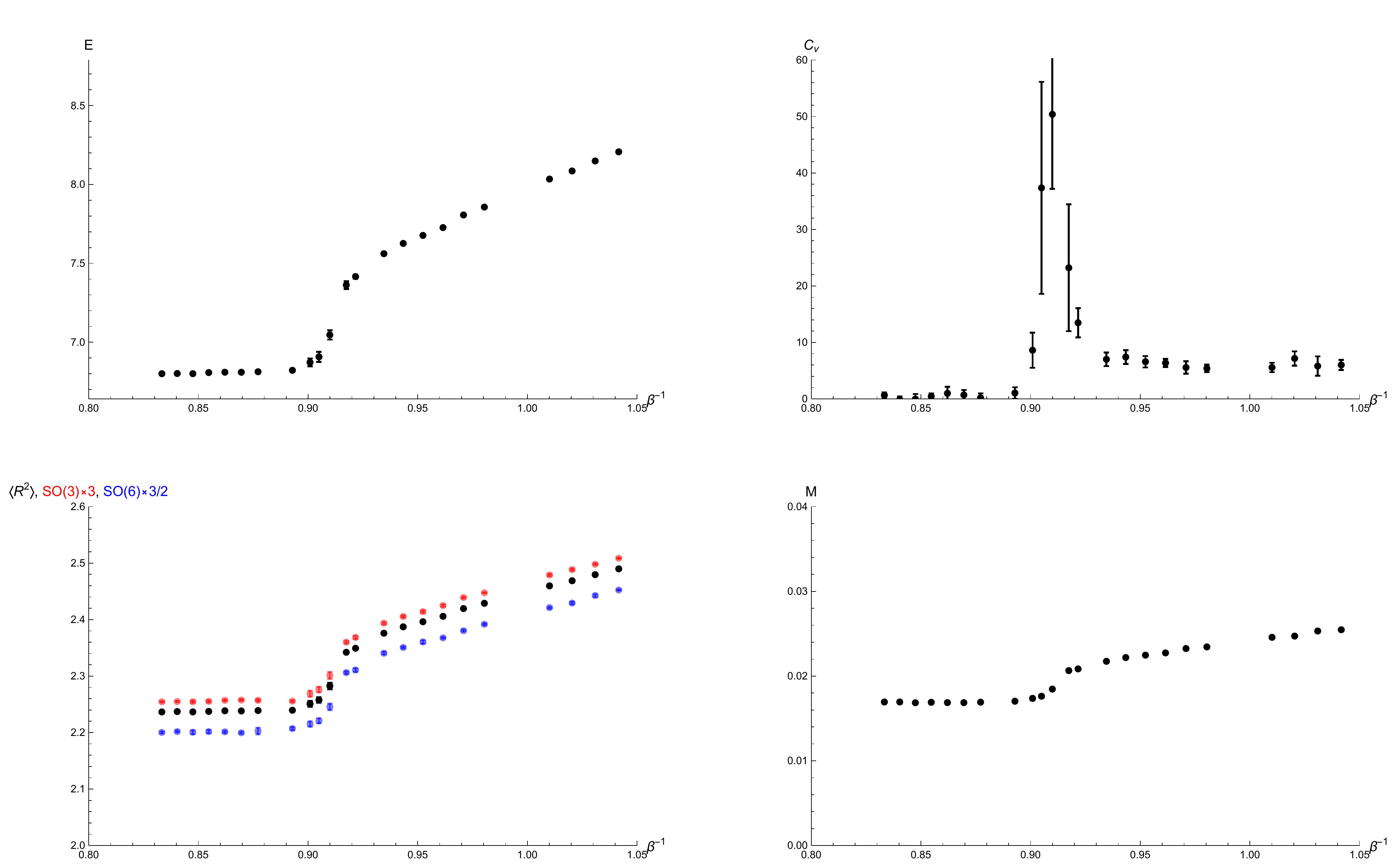}
\caption{\small Observables of the model for $\mu = 2,\ N = 32\mbox{
    and } \Lambda = 24$. The Myers observable seems to be negligible and copies the shape of $\langle R^2\rangle$. All observables point to either a single or multiple
  transitions around $T \approx 0.92$. The split between $SO(3)$ and $SO(6)$ components of $\langle R^2\rangle$ is due to different masses.}
\label{Fig:Observables}
\end{figure}

\begin{figure}[t] 
\includegraphics[width=6in]{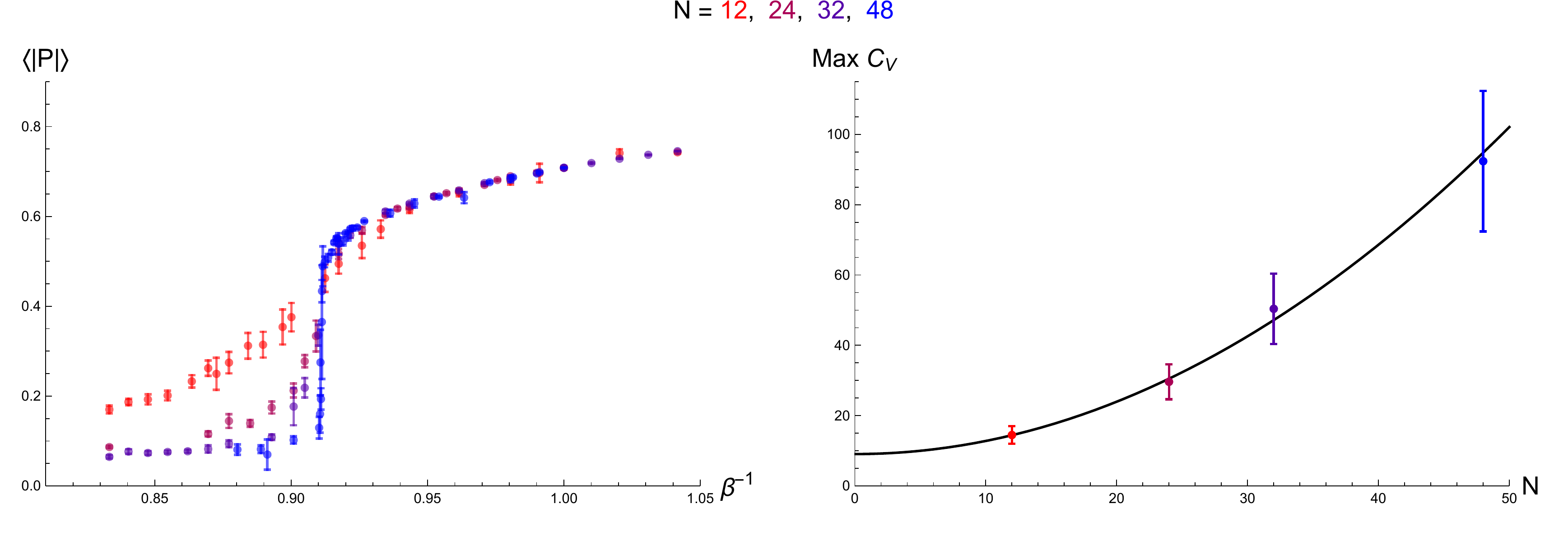}
\caption{\small
  The evolution of the Polyakov loop $\langle|P|\rangle$ for $\mu = 2,\ \Lambda = 24$ with increasing $N$. The transition region becomes sharper with larger $N$. The right figure grows as $C_{V}^{\mbox{max}} = 9.1(8) + 0.037(2)N^2$. }
\label{Fig:pols}
\end{figure}

The figure \ref{Fig:Observables} shows the behaviour of the observables for $\mu = 2, \ N =32 \mbox{ and } \Lambda = 24$. We can clearly see that the system undergoes either one or more (closely separated) phase transitions around $T \sim 0.92$. Measurements of the Polyakov loop and specific heat for increasing values of $N$ are shown in figure \ref{Fig:pols}, confirming that the transition region shrinks in the large-$N$ limit and the scaling of $C_{V}^{\mbox{max}}$ signals a 1st order phase transition.

\begin{figure}[t] 
  \includegraphics[width=6in]{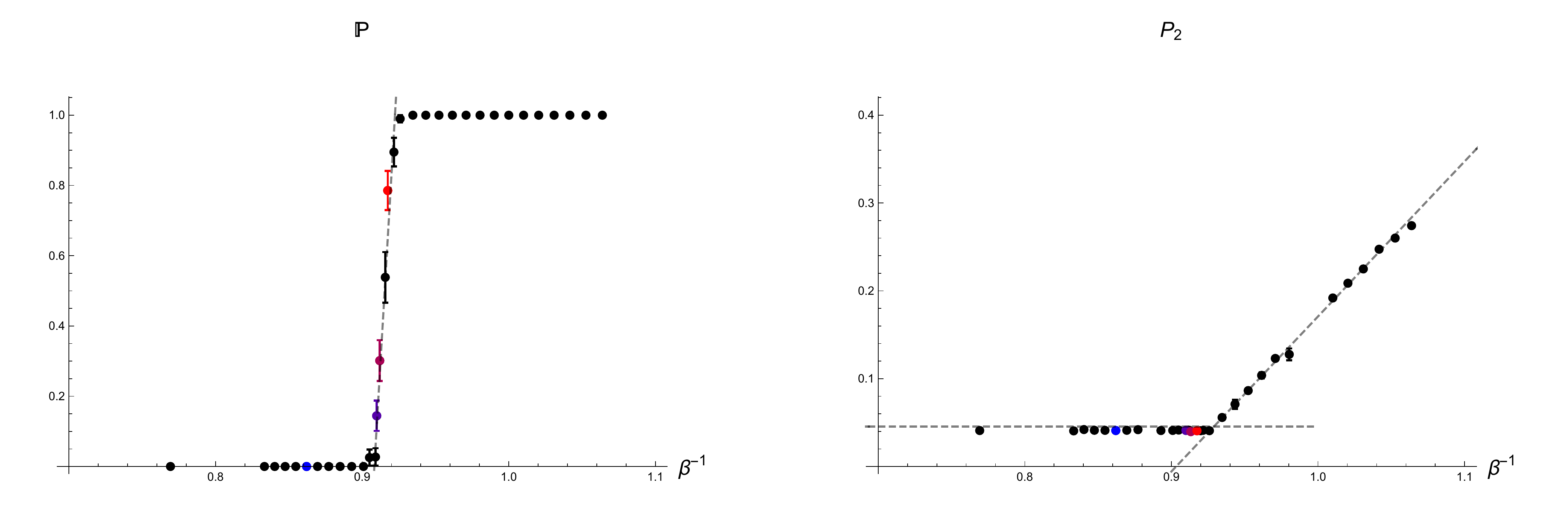}
  \caption{\small Values of $\mathbb{P}$ and $\langle\vert P_2\vert\rangle$
    for $\mu = 2,\ N = 32\mbox{ and } \Lambda = 24$ and increasing value of temperature $T = \beta^{-1}$. The points in the transition region in the left plot
  were fit by a linear function whose slope increases with
   $N$. The four coloured points correspond to $T = 0.8621,\ 0.9099,\ 0.9120 \mbox{ and } 0.9174$.}
\label{Fig:Q}
\end{figure}

Let us now focus on the case of $\mu = 2, \ N =
  32\mbox{ and } \Lambda = 24$. The root of the growing linear function in the left panel of figure \ref{Fig:Q} is taken to be at the first (pseudo)critical temperature, $T_{c1}$, where the underlying eigenvalue distribution becomes nonuniform. The bending point in the function  in the right panel, which is measured as a crossing point of two linear fits, is taken to be at the second (pseudo)critical temperature, $T_{c2}$, where the eigenvalue distribution becomes gapped. Details of this behaviour are discussed in \cite{Asano:2020yry}. 

\begin{figure}[t] 
\includegraphics[width=6in]{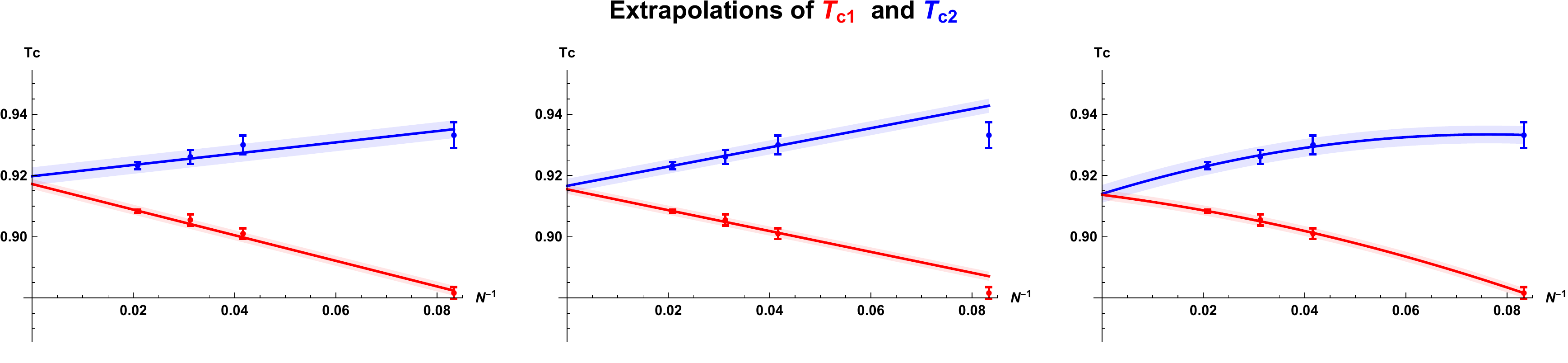}
\caption{\small
Extrapolations of the (pseudo)critical temperatures from results obtained from $N = 12,\ 24,\ 32,\ 48$. The left plot is using a linear fit, the middle plot is using linear fit while omitting $N=12$ value, the right one is using a quadratic fit. All fits are performed functions of $N^{-1}$.}
\label{Fig:extrap}
\end{figure}

We have measured the values of $T_{c1}$ and $T_{c2}$ for $N = 12,\ 24,\ 32,\ 48$ and extrapolated them to infinite $N$, the results are shown in the figure \ref{Fig:extrap}. The (pseudo)critical temperatures merge into a single one, the exact value depends only slightly on the choice of the fitting function. The best agreement is for the quadratic fit, yielding $T_{c1} \rightarrow 0.9137(9)$ and $T_{c2} \rightarrow 0.914(2)$ in the large-$N$ limit.

\begin{figure}[t] 
  \includegraphics[width=6in]{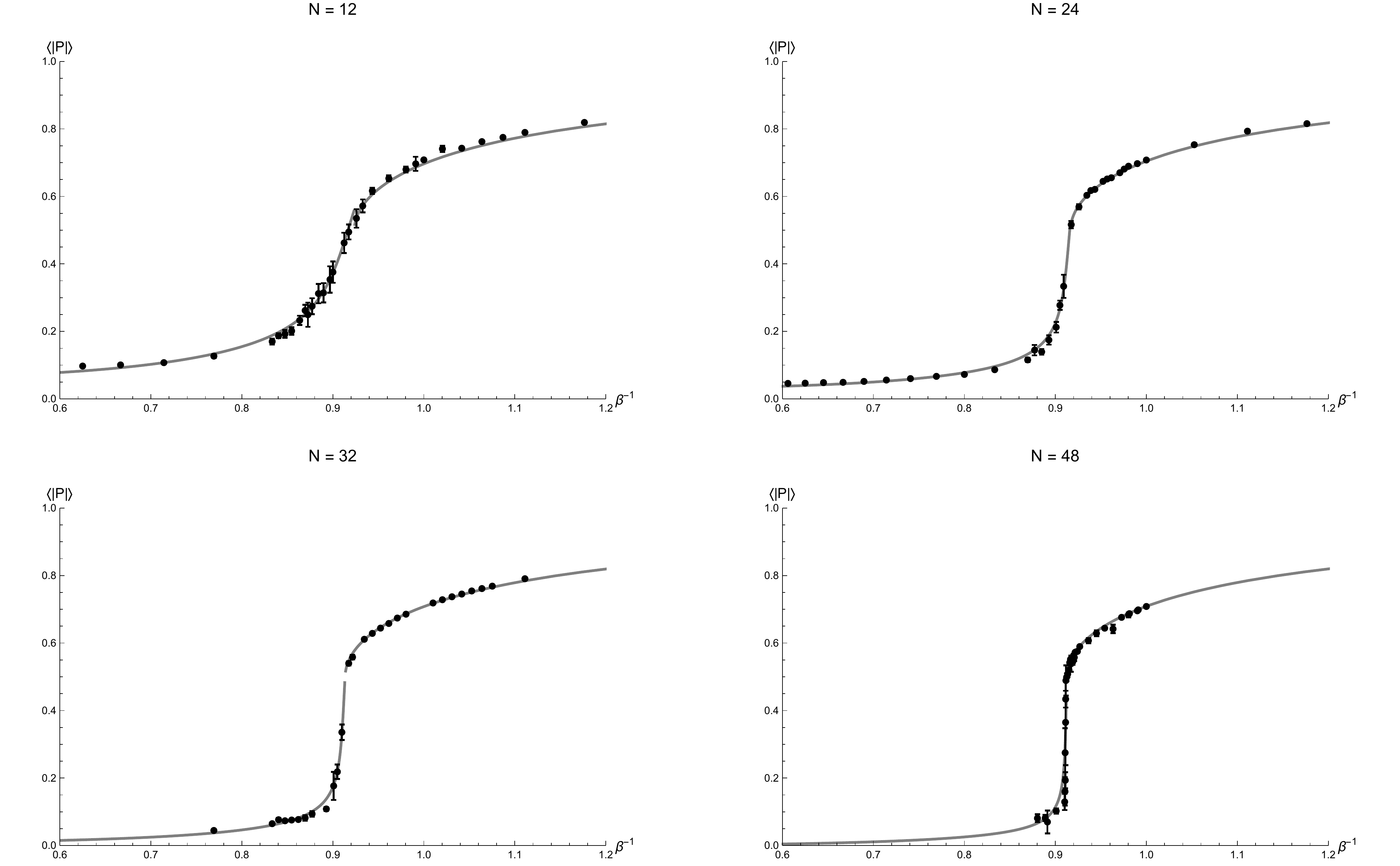}
\caption{\small Numerically obtained values of $\left\langle|P|\right\rangle$ for $\mu = 2, \Lambda = 24$ and various values of $N$. The solid lines are fits using theoretical predictions \ref{1oNfit1} and \ref{1oNfit2} with parameters $T_H \mbox{ and } P_0$ obtained by fitting. }
\label{Fig:1oNfits}
\end{figure}

The critical temperature can be approximately obtained even with a single, possibly small, value of $N$. To do so, one needs to have a good theoretical prediction for $\langle|P|\rangle$ as a function of $T$ with finite-$N$ corrections included. We used

\begin{equation} \label{1oNfit1}
\left\langle|P|\right\rangle (T) = P_0 +\frac{ \sqrt{\langle l\rangle_N}}{N}e^{ -m \left(T^{-1}-T_H^{-1}\right)} \mbox{ for } T<T_H
\end{equation}

\begin{equation}\label{1oNfit2}
\left\langle|P|\right\rangle (T) = \frac{1}{2} \frac{e^{m \left(T^{-1}-T_H^{-1}\right)}}{1-\sqrt{1-e^{m(T^{-1}-T_H^{-1})}}} \mbox{ for } T>T_H.
\end{equation}
Here, $\langle l\rangle_N = \frac{e^{m\left(T^{-1}_H-T^{-1}\right)}}{1-e^{m\left(T^{-1}_H-T^{-1}\right)}}-\frac{c\ e^{\left(T_H^{-1} - T^{-1}\right) c\ m\ N^2} N^2}{1-e^{\left(T_H^{-1} - T^{-1}\right) c\ m\ N^2}}$, $m = T_H \ \ln 9$ and $c$ is chosen so the two functions meet at $T=T_H$. This is obtained in the Hamiltonian approach to the gauge Gaussian model, \cite{Furuuchi:2003sy, Hadizadeh:2004bf} and will be discussed in our forthcoming work \cite{Asano:2020pre}. $T_H$ is to be interpreted as the Hagedorn temperature and we have measured its values for $N=12,\ 24,\ 32,\ 48$ as $T_H = 0.924(1),\ 0.9167(4)$, $ 0.9136(3),\ 0.9127(2)$. This is very close to the results obtained from the previous method of extrapolating two (pseudo)critical temperatures. The values of $P_0$ are zero-temperature contributions to $\langle|P|\rangle$ and are understood to be only finite-$N$ effects, their values for $N=12,\ 24,\ 32$ are $P_0 = 0.058(4),\ 0.028(2),\ 0.008(3)$. We have set $P_0=0$ for $N=48$ as we did not obtain enough data points for $T\ll T_H$. The results are shown in the figure \ref{Fig:1oNfits}. 

\begin{wrapfigure}{l}{0.6\textwidth}
  \includegraphics[width=0.9\linewidth]{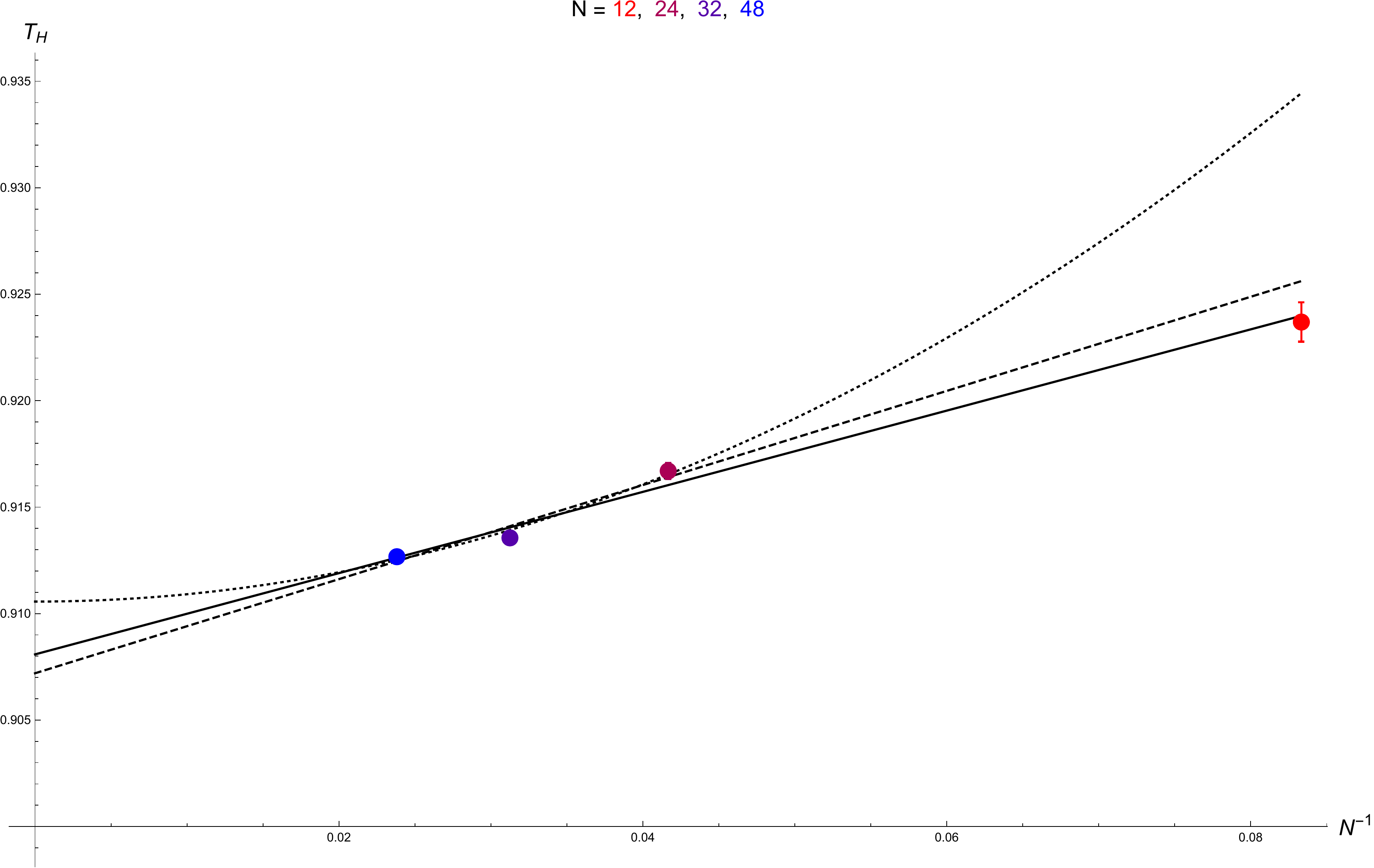}
\caption{\small Extrapolations of the Hagedorn temperatures $T_H$ for $N = 12, \ 24, \ 32, \ 48$. The lines correspond to various fitting functions: linear (solid), linear with $N=12$ point omitted (dashed), quadratic $a + b N^{-2}$ with $N=12$ omitted (dotted). The large-$N$ values are, in the same order, $T_H = 0.9081(6), \ 0.907(2), \ 0.9106(6)$. The errorbars are the fitting errors of $T_H$ and seem to be slightly underestimated.}
\label{Fig:TH_extrap}
\end{wrapfigure}

We can take $T_H$ for various values of $N$ and extrapolate them to the large-$N$ limit, various choices of the fitting function are shown in the figure \ref{Fig:TH_extrap}. The fitting function $a + b N^{-2}$ gives $T_H = 0.9106(6)$ in this limit, an estimate reasonably close to the merger point of the two (pseudo)critical temperatures in the same limit.

The two methods described above can be applied to the model for any value of $\mu$. At $\mu=0$ the model is just the bosonic part of the BFSS model which has been well researched both theoretically and numerically. At first, it was believed that there are two, closely separated phase transitions (second order and third order). The latest research \cite{Bergner:2019rca}, however, reports only a single 1st order phase transition. Our $\mu=2$ extrapolations to infinite $N$ are in an agreement with a single phase transition of the Hagedorn type, i.e.~a 1st order transition with strong finite $N$ effects in the confined phase (see the figure \ref{Fig:1oNfits}). 

\begin{wrapfigure}{l}{0.6\textwidth}
  \includegraphics[width=0.9\linewidth]{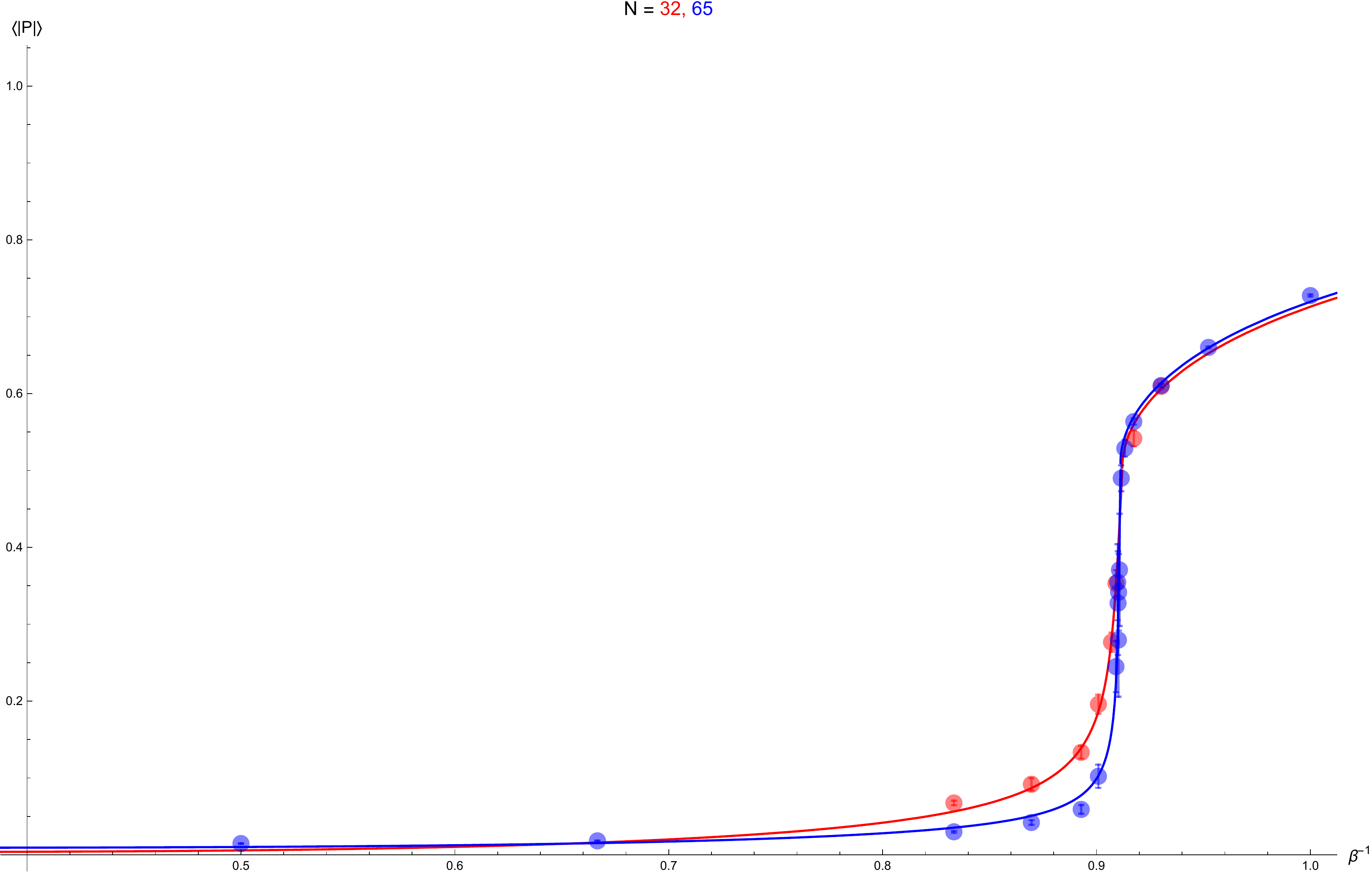}
\caption{\small We have tested the theoretical predictions \ref{1oNfit1} and \ref{1oNfit2} also for the pure Gaussian model with $\mu=2$ for which the critical temperature is known to be exactly $1/ln(3)$. In these plots we have set $P_0=0$.}
\label{Fig:1oNfits2}
\end{wrapfigure}

We have performed a detailed study of the gauge Gaussian model with $\mu=2$ and shown that the leading finite-$N$ effects in the low temperature phase are substantial. The results are shown in the figure \ref{Fig:1oNfits2}. The solid curves are those described by $\langle l\rangle_N$ discussed above where $\langle l\rangle_N$ uses the sharp cutoff on states in the Hamiltonian formulation described by words of maximum length $c N^2-1$ and $\langle l\rangle_N$ describes the mean word length (see \cite{Furuuchi:2003sy, Hadizadeh:2004bf}).

For large values of $\mu$, only the quadratic terms contribute and the model effectively reduces to a gauged Gaussian model that has a single critical temperature located at $T_c = \frac{\mu}{6 \ln \left(3 + 2 \sqrt{3}\right)}$.

\begin{wrapfigure}{l}{0.6\textwidth}
  \includegraphics[width=0.9\linewidth]{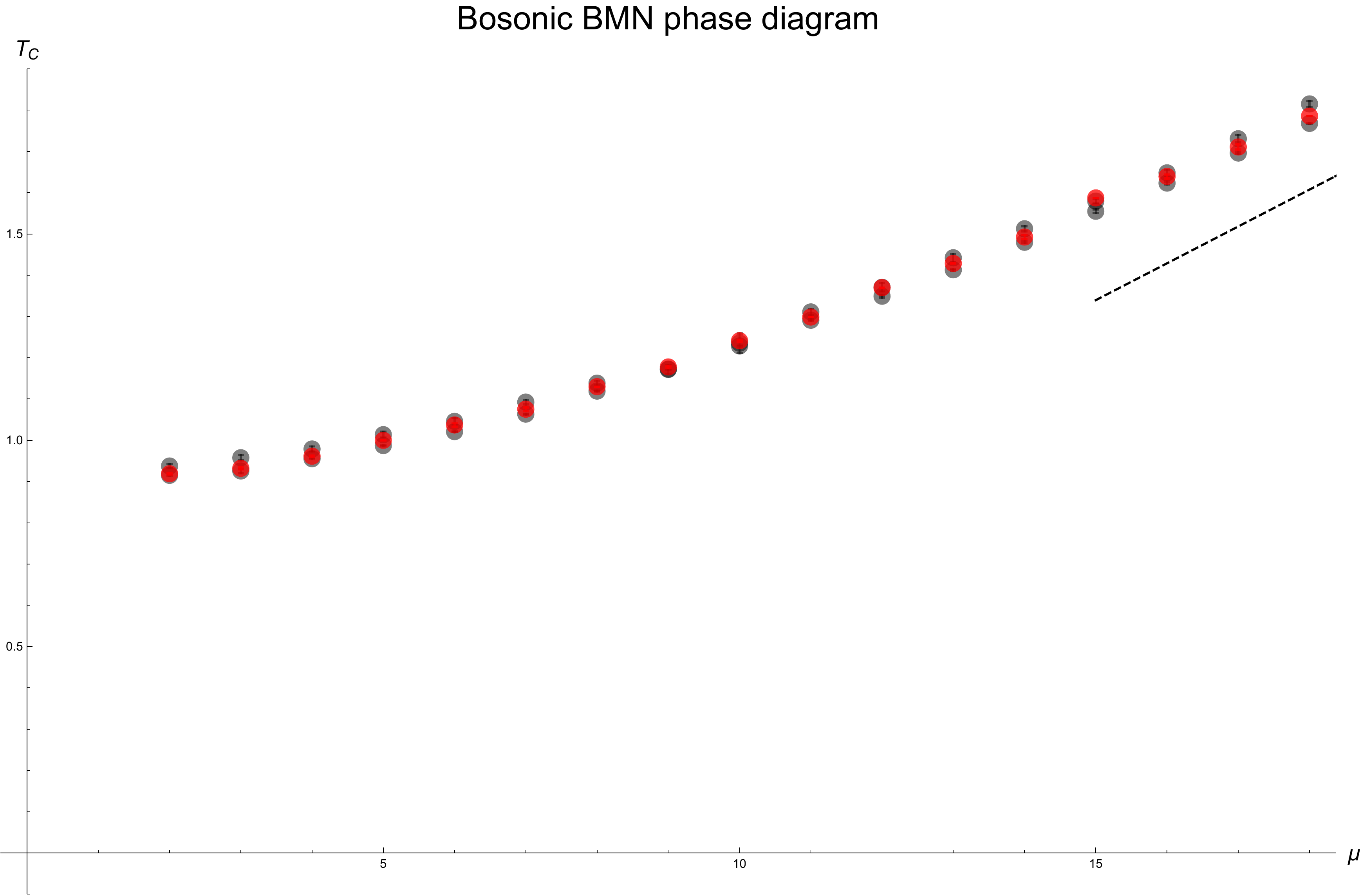}
\caption{\small The (pseudo)critical temperatures for $N=\Lambda=24$. The gray points were obtained using $\mathbb{P}$ and  $\langle\vert P_2\vert\rangle$. The red points were obtained by measuring $T_H$ using \ref{1oNfit1} and \ref{1oNfit2}. The dashed line shows the large-$\mu$ prediction $T_c = \frac{\mu}{6 \ln \left(3 + 2 \sqrt{3}\right)}$.}
\label{Fig:Phase diagram}
\end{wrapfigure}

We have produced the phase diagram for $N=\Lambda=24$ which is shown in the figure \ref{Fig:Phase diagram}. The gray points mark two (pseudo)critical temperatures measured using $\langle\vert P_2\vert\rangle$ and $\mathbb{P}$. The red points show the critical temperature measured by $T_H$, which, as expected, lies between the other two. The dashed line shows the large-$\mu$ critical temperature which the points asymptote to. 

\section{Conclusions}

We have analysed the behaviour of the bosonic BMN matrix model, focusing on the thermal deconfining phase transition at finite $\mu$. We observed that at finite $N$ we can distinguish two closely separated (pseudo)critical temperatures $T_{c1}$ and $T_{c2}$. At $T>T_{c1}$ the system starts to stay in the state with $\langle|P|\rangle\sim 1/2$ and the first moment of the eigenvalue distribution begins its increase. At $T>T_{c2}$ higher moments of the eigenvalue distribution develop nontrivial expectation values, so that the distribution is gapped. We have also observed that these two (pseudo)critical temperatures merge into one in the large-$N$ limit. 

We were able to fit the data for the Polyakov loop $\langle|P|\rangle$ using functions obtained from a theoretical description of the model at finite $N$. The fitting parameter $T_H$ is to be interpreted as the Hagedorn temperature (details will be discussed in our upcoming work), which is consistent with the aforementioned single large-$N$ critical temperature. 

The two detailed methods used in this paper for $\mu=2$ give consistent estimates for the critical temperature in the large-$N$ limit. Combining those two yields the value $T_c = 0.912(2)$. 

The exact nature of the phase transition remains unclear at this point. Analysing the Monte Carlo trajectories of the system shows clear signs of two-level system well approximated by two Gaussian distributions \cite{Fisher:1982xt,Landau:1984}. However, fitting using \ref{1oNfit1} and \ref{1oNfit2} shows a clear relation to the Hagedorn phase transition as well. 

For a single finite value of $N$, we have constructed the phase diagram, which interpolates smoothly between the zero-mass BFSS prediction and large-mass prediction of the gauged Gaussian model. Numerical simulations strongly suggest that the two (pseudo)critical temperatures shown in the figure \ref{Fig:Phase diagram} merge into one in the large-$N$ limit, possibly close to the value predicted by the Hagedorn fit (red points in the same diagram). In \cite{Asano:2020yry} we have also tested that with our lattice formulation the results depend only very weakly on lattice parameter $\Lambda$ and are reasonably close to the continuum value. 

Our choice of the fitting function, equations \ref{1oNfit1} and \ref{1oNfit2}, contained a contribution from the $T \rightarrow 0$ behaviour of the Polyakov loop, denoted by $P_0$. We know that in this limit the eigenvalues of $A$ are uniformly distributed over the entire interval. We can model them as a set of random numbers with Gaussian distribution with mean values $\mu_j=\frac{2 \pi j}{N}$ and standard deviation $\sigma$. This way, $N$ and $\sigma$ determine the value of $\langle|P|\rangle$.

For $N=12,\ 24,\ 32$ we have obtained, for the data at the lowest measured temperatures ($\beta = 2.2,\ 1.85,\ 1.7$) the values of $\sigma$ and used it to compute $P_0$. The results of the calculation ($0.082,\ 0.043,\ 0.029$) are very close to the values of $\langle|P|\rangle$ measured at those temperatures  ($0.08(2),\  0.0427(9),\ 0.034(1)$). Given the knowledge of $\sigma$, we can estimate the value of $P_0$ rather precisely.

The next step is, instead of measuring $\sigma$, to have a theoretical estimate for it. The dominant effect in the zero-temperature limit is the logarithmic repulsion between the eigenvalues. We can estimate $\sigma$ by assuming all but one eigenvalues to be fixed at $\theta_j = \frac{2 \pi j}{N}$. Then, we expand the potential in terms of the unfixed eigenvalue and use the coefficient in the quadratic term to estimate the typical value of $\sigma$.  This yields, given the estimated values of $\sigma$, the estimate for $\langle|P|\rangle$ as ($0.083,\ 0.029,\ 0.019$) which, given the bold estimates, is reasonably close to the measured values. Therefore, we believe that describing the low temperature behaviour of the gauge field using a set of uniformly separated eigenvalues fluctuating around their mean positions in the presence of logarithmic repulsion is accurate. 

Our results are in broad agreement with $1/D$ studies \cite{Mandal:2009vz, Takeuchi:2017wii} but do not match it exactly as the authors observe two closely separated phase transitions. As a recent numerical study of the BFSS model \cite{Bergner:2019rca} also reports a single phase transition, we believe that by including higher terms in the $1/D$ expansion, the two phase transitions would merge into a single one in this approximation as well. 

A possible line of future research is the study of bosonic version of the D0--D4
Berkooz-Douglas model \cite{Filev:2015cmz,Asano:2016xsf,Asano:2016kxo}. The model has degrees of freedom that transform under the fundamental representation of $SU(N_f)$ and the work \cite{Asano:2018nol} reported exceptional behaviour for $N_f = 2 N$ which should be interesting to study in the bosonic model. 

\subsection*{Acknowledgment}
The authors wish to thank the Corfu Summer Institute for its hospitality and acknowledge the Irish Centre for High-End Computing (ICHEC) for the provision of computational facilities and
support (Projects dsphy009c, dsphy010c and dsphy012c). The support from
Action MP1405 QSPACE of the COST foundation is gratefully
acknowledged.  Y.~Asano is supported by the JSPS Research Fellowship
for Young Scientists. S. Kov\'a\v{c}ik was supported by Irish Research
Council funding. The authors would like to thank G.~Bergner,
M.~Hanada, G.~Ishiki, T.~Morita and H.~Watanabe for valuable discussion.

\end{document}